\documentclass[lettersize,journal]{IEEEtran}
\usepackage{amsmath,amsfonts}
\usepackage{algorithmic}
\usepackage{algorithm}
\usepackage{array}
\usepackage[utf8]{inputenc}
\usepackage{textcomp}
\usepackage{url}
\usepackage{xcolor}
\usepackage{verbatim}
\usepackage[utf8]{inputenc}
\usepackage{graphicx}
\usepackage{multirow}
\usepackage{multicol}
\usepackage{caption}
\usepackage{subcaption}
\usepackage[colorlinks=true,
            linkcolor=blue,
            citecolor=blue,
            urlcolor=blue]{hyperref}
\usepackage{cite}
\usepackage{doi}
\usepackage{booktabs}
\usepackage{booktabs,siunitx,adjustbox}
\begin{document}

\title{3D Wi-Fi Signal Measurement in Realistic Digital Twin Testbed Environments Using Ray Tracing}

\author{Mengyuan Wang,
Haopeng Wang,
Haiwei Dong, \IEEEmembership{Senior Member, IEEE,}
Abdulmotaleb El Saddik, \IEEEmembership{Fellow, IEEE}
\thanks{Mengyuan Wang, Haopeng Wang, and Abdulmotaleb El Saddik are with the School of Electrical Engineering and Computer Science, University of Ottawa, Ottawa, ON K1N 1A2, Canada (E-mail: \{mwang259; hwang266; elsaddik\}@uottawa.ca).}
\thanks{Haiwei Dong is with University of Ottawa and Huawei Canada, Ottawa, ON K2K 3J1, Canada (E-mail: haiwei.dong@ieee.org).}
\thanks{Corresponding author: Haiwei Dong.}}

\markboth{IEEE Transactions on Instrumentation and Measurement}%
{Shell \MakeLowercase{\textit{et al.}}: A Sample Article Using IEEEtran.cls for IEEE Journals}


\maketitle

\begin{abstract}

Accurate and efficient modeling of indoor wireless signal propagation is crucial for the deployment of next-generation Wi-Fi. This paper presents a digital twin-based measurement system that integrates real-world 3D environment reconstruction with deterministic ray tracing for physically grounded electromagnetic modeling. Building geometry is obtained through LiDAR scanning, followed by object segmentation and assignment of ITU-R standard material parameters. The propagation process is simulated with a GPU-accelerated ray-tracing engine that generates path-level channel attributes, including delay, power, angular dispersion, and Ricean K-factor. Under identical runtime constraints, the proposed system is evaluated against a commercial measurement simulator, demonstrating up to 21 dB higher path gain and consistently improved signal-to-interference-plus-noise ratio in line-of-sight conditions. Additionally, experiments against onsite RSSI measurements confirm a high spatial correlation of 0.98 after calibration, proving the system's fidelity in real-world settings. Furthermore, coverage analysis across 2.4 GHz, 5 GHz, and 6 GHz bands demonstrates the capability of system to model frequency-dependent material attenuation for Wi-Fi 6E/7 networks. Finally, the system offers interactive 3D visualization and on-demand data extraction, highlighting its potential for digital twin-driven wireless system design and optimization.
\end{abstract}


\section{Introduction}
Wireless communication has become indispensable in many indoor spaces, such as homes, offices, and factories. Among the various technologies, Wi-Fi, based on the IEEE 802.11 standards, is one of the most widely deployed due to its low cost and ease of installation. As networks evolve toward Wi-Fi 6E/7, the demand for accurate characterization of signal behavior in complex indoor environments has become increasingly critical.
In the deployment of indoor wireless networks, accurate and efficient signal measurement is essential for evaluating coverage, optimizing system performance, and guiding network planning. Traditional measurement methods typically rely on technicians carrying specialized equipment into the environment and conducting repeated field measurement. This process is time-consuming, labor-intensive, and difficult to reproduce consistently under different conditions, which makes large-scale or rapid channel characterization impractical.

To overcome these challenges, researchers turn to simulation-based methods that can generate results matching or even exceeding those of existed measurement tools. Some measurement tools use simplified or synthetic indoor layouts, while others cannot properly model how different materials affect signal strength. Moreover, few tools allow users to import real 3D spaces or assign realistic electromagnetic properties to objects. Recent advances in ray tracing have leveraged differentiable propagation modeling to enhance simulation flexibility \cite{RT}, but challenges remain in integrating real-world data. These limitations make it difficult to perform precise network planning, which typically involves optimizing access point placement, channel allocation, and interference management to ensure reliable coverage and sufficient capacity in real-world deployments.

Commercial tools such as Wireless InSite \cite{remcom2024} and WinProp \cite{winprop2024} support high-quality simulations but are costly, closed-source, and hard to customize. On the other hand, most open-source tools either work with basic geometries or lack support for advanced material modeling. GPU-accelerated ray tracing engine has been utilized for path loss prediction in urban environments using accurate scene models \cite{Path-Loss-Prediction}, addressing some of these gaps with GPU acceleration. There is a need for a simulation system that can handle high fidelity virtual environments, support material-aware modeling, and remain flexible and reproducible for research/industry use.

In this paper, we propose a digital twin-based simulation system that models indoor wireless propagation using ray tracing. We reconstruct real 3D environments using LiDAR scans, segment objects like walls and furniture, and assign material properties based on International Telecommunication Union (ITU) standards. These environments are then used with the ray tracing engine to simulate signal paths and compute key metrics such as delay, path gain, and Signal-to-Interference-plus-Noise Ratio (SINR). The main contributions of our work include:
\begin{itemize}
    \item To the best of our knowledge, this paper introduces the first measurement system that integrates real-world scans, object segmentation, and standardized material modeling to construct digital twin environments.     
    \item We propose a ray tracing method to simulate signal propagation  which supports programmable, GPU-accelerated ray tracing and outputs path-level channel data.
    \item Our system is able to extract key channel attributes that are typically obtained through hardware-based measurements, thus providing a new measurement approach for wireless signal research.
    \item We conduct a deep comparison experiment to evaluate proposed system with the commercial tool Wireless InSite under the same simulation time.
\end{itemize}

\section{Related Work}
Accurate wireless channel modeling is critical for evaluating communication protocols, deployment strategies, and resource allocation. Traditional approaches often rely on statistical models, such as ITU-R P.1238 \cite{ITU-P1238} and COST 231 \cite{COST231LTE}, which estimate large-scale path loss using empirical regression. These models are computationally efficient and well-suited for outdoor or simple indoor environments with coarse scene labels (e.g., “office”, “corridor”). However, they fail to capture structural variations such as wall materials or occlusions, facing challenges for high-frequency networks requiring fine-grained spatial resolution in complex indoor settings. Deterministic methods offer greater physical accuracy by numerically solving Maxwell’s equations. Full-wave solvers like FDTD \cite{FDTD} and FEM \cite{FEM} excel in modeling detailed field behavior, making them valuable for small-scale, high-precision applications such as antenna design. Nevertheless, their high computational cost and stringent mesh resolution requirements render them impractical for large-scale indoor simulations, preventing the adaption as their practical measurement instruments.

Ray Tracing (RT) strikes a balance between accuracy and efficiency by modeling signal propagation as geometric rays interacting with the environment through reflection, diffraction, and transmission. Commercial tools such as Wireless InSite \cite{remcom2024} and WinProp \cite{winprop2024} are widely adopted for their high simulation fidelity. However, their closed-source nature, license restrictions, and limited automation capabilities hinder reproducibility and large-scale experimentation. Recent efforts have explored RT for robust indoor channel modeling at 2.4 GHz and 5 GHz frequencies \cite{Robust-Channel-Modeling}, integrating empirical measurements to address multipath effects in complex environments, thus motivating the need for configurable and customizable system solutions.

To address these limitations, the ray tracing engine module \cite{sionna} enables GPU-accelerated path-level simulations within programmable environments. It supports realistic 3D scene imports, material property configurations, and output of detailed metrics such as delay, gain, and phase for channel impulse response (CIR) construction and signal coverage analysis. Ray tracing techniques have been applied to improve signal modeling in challenging environments \cite{Ray-Tracing-GNSS}, motivating our research of RT for Wi-Fi propagation simulation.

Nonetheless, most prior RT-based studies rely on synthetic or manually constructed scenes, lacking integration with real-world spatial data and standardized material modeling. Previous work has utilized 3D ray tracing for indoor environment modeling \cite{RT-Fingerprint-Based-Indoor-Position}, providing a foundation for our LiDAR-based digital twin reconstruction process. Ray tracing has been explored to enhance indoor positioning by integrating CIR measurements \cite{Ray-Tracing-Assisted-Fingerprinting-cir}, offering valuable insights into multipath modeling that inspire digital twin approach. In this work, we bridge this gap by proposing a digital twin-based RT simulation system that combines LiDAR-based reconstruction, object segmentation, and ITU-compliant material assignment, enabling reproducible and physically grounded channel modeling in complex indoor environments. These limitations motivate the development that not only models propagation accurately but also functions as a flexible measurement tool for indoor wireless research.

\section{Wireless Propagation Simulation Method}
\begin{figure*}[htbp]
\centerline{\includegraphics[width=43pc]{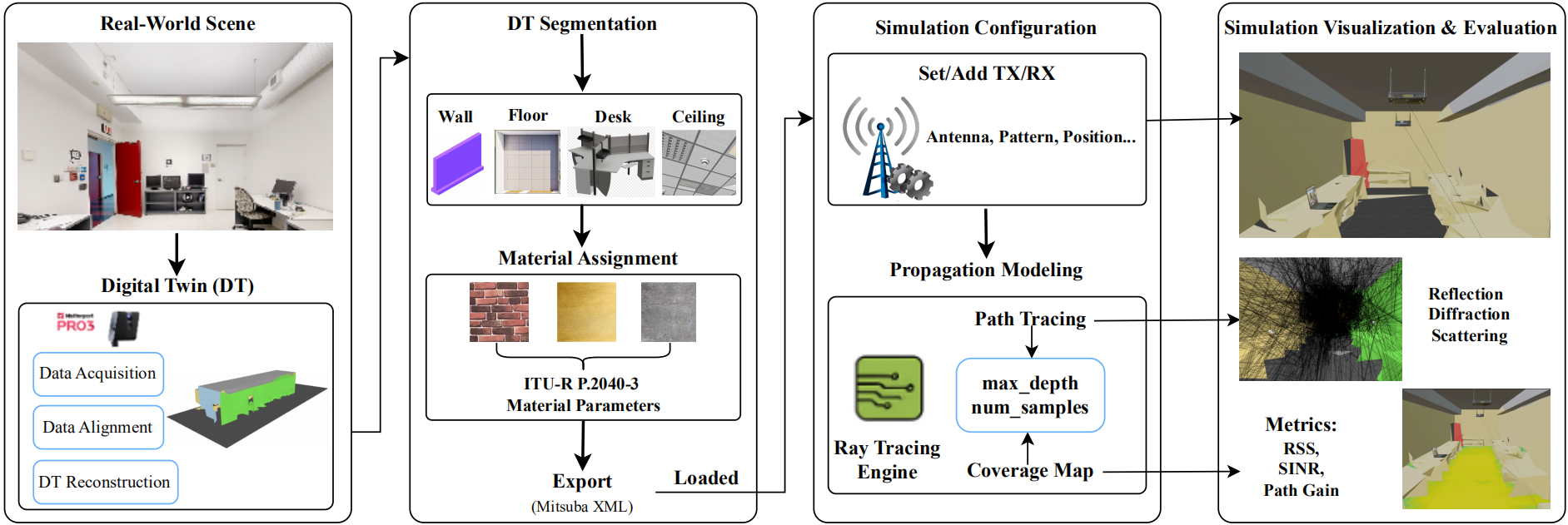}}
\caption{The architecture of the proposed wireless propagation simulation. It begins with capturing real-world environments using a LiDAR-based scanning camera, followed by digital twin reconstruction. The electromagnetic materials are assigned to segmentated objects based on ITU-R P.2040-3 parameters, then, scene loaded into the simulation pipeline, where transmitters and receivers are added along with configurable parameters. Using a GPU-accelerated ray-tracing engine, both per-path channel characteristics and coverage metrics are computed and visualized.}
\label{sionna architecture}
\end{figure*}

We develop a measurement system that integrates digital twin reconstruction with physics-based ray tracing for indoor wireless signal modeling. As illustrated in Fig. \ref{sionna architecture}, the system comprises four functional modules: real-world scene acquisition, object segmentation with material assignment, simulation configuration, and visual evaluation. These modules collectively support the generation of high-resolution CIR and performance maps under diverse transmitter-receiver arrangements.

\subsection{Digital Twin Reconstruction} 
The 3D building environment was generated using the Matterport Pro3 scanning device to create a high-precision digital twin suitable for signal propagation simulation. The built-in LiDAR sensor operates at a wavelength of 904 nm and provides a $360^\circ$ horizontal and $295^\circ$ vertical field of view. It functions within a minimum operating range of 0.5 meters which extends to 100 meters when utilizing raw E57 data export. The sensor maintains a ranging accuracy of $\pm 20$ mm at a reference distance of 10 meters. Each scan acquires 1.5 million depth points at a rate of 100,000 points per second. The photographic subsystem captures five exposures per frame to generate a high dynamic range image. This process produces a final panorama with a resolution of 134.2 megapixels. The system completes a full rotational capture in less than 20 seconds. These specifications ensure that the reconstructed geometry meets the precision requirements for ray tracing analysis. The process began with data acquisition across a six-story building with a total area of 17,567 square meters. The campaign involved 1,099 distinct scanning points. We placed visual markers at intervals throughout the building to ensure comprehensive data capture and minimize alignment errors. These markers acted as reference points for aligning scans and mitigating errors caused by environmental complexities such as repetitive patterns or dynamic elements. The device utilized its LiDAR sensor to generate point cloud data and a high-resolution camera to capture RGB images, providing both geometric and visual detail \cite{lidarscanning}.

The captured data underwent a two-stage alignment process to ensure precision. Intra-scan alignment integrated RGB images with the LiDAR point cloud by mapping the color information onto the 3D geometric data. This step ensured a textured representation of the environment with precise correspondence between visual and spatial features. In details, each LiDAR point $\mathbf{x}_i=(x_i,y_i,z_i)$ is projected onto the RGB image plane using the calibrated intrinsic and extrinsic camera parameters
\begin{equation}
\mathbf{u}_i = \mathbf{K}\big[\mathbf{R}\mathbf{x}_i + \mathbf{t}\big]
\end{equation}
where $\mathbf{K}$ is the intrinsic camera matrix, and $(\mathbf{R},\mathbf{t})$ represent the rotation and translation between the LiDAR and camera. The pixel value at $\mathbf{u}_i$ is then assigned to the 3D point, forming a colored point cloud $\mathcal{P}=\{(\mathbf{x}_i,\mathbf{c}_i)\}_{i=1}^N$.

Inter-scan alignment was then employed to merge individual scans into a unified 3D coordinate system. This process involved detecting salient features across scans and applying the Iterative Closest Point (ICP) algorithm to minimize the alignment error
\begin{equation}
E(\mathbf{R},\mathbf{t}) = \sum_{i=1}^M \left\| \mathbf{R}\mathbf{x}_i + \mathbf{t} - \mathbf{y}_{\pi(i)} \right\|^2 
\end{equation}
where $\{\mathbf{x}_i\}$ are source points, $\{\mathbf{y}_{\pi(i)}\}$ are their matched target points, and $(\mathbf{R},\mathbf{t})$ is iteratively optimized to achieve global consistency. This approach significantly reduced positional drift and ensured that the entire six-floor model was spatially coherent.

To reconstruct the final model, the aligned point cloud data was processed in the cloud using the Advancing Front (AF) algorithm. AF is a surface reconstruction method widely used in computational geometry and mesh generation. Unlike Poisson reconstruction, AF does not impose watertightness constraints, making it particularly suitable for architectural structures with open boundaries, such as corridors, windows, and doors. The algorithm begins from an initial front (set of edges or triangles) and iteratively advances by introducing new vertices and faces. At iteration $k$, a front edge $(v_i,v_j)$ is selected, and a new vertex $v_k$ is generated by optimizing a quality metric $\phi$
\begin{equation}
v_k = \arg\min_{v \in \Omega} \, \phi(v; v_i,v_j)
\end{equation}
where $\Omega$ is the feasible region defined by the local sampling density, and $\phi(\cdot)$ enforces geometric quality (e.g., minimum angle or aspect ratio). 

An important property of AF is its adaptive resolution capability. The element size $h(\mathbf{x})$ is adjusted according to the local point density $\rho(\mathbf{x})$
\begin{equation}
h(\mathbf{x}) \propto \frac{1}{\sqrt{\rho(\mathbf{x})}}
\end{equation}
ensuring that densely sampled areas (e.g., glass walls, staircases) are reconstructed with high fidelity, while sparsely sampled regions are represented more coarsely to improve efficiency.

To handle the large scale of the scanned building, a block-wise parallel processing strategy was adopted. The point cloud was partitioned into blocks $\{\mathcal{P}_b\}$ based on spatial coordinates, each reconstructed independently using AF. Overlapping regions between adjacent blocks were aligned and merged through local ICP refinements
\begin{equation}
\mathcal{M} = \bigcup_b \text{AF}(\mathcal{P}_b)
\end{equation}
where $\mathcal{M}$ represents the final integrated mesh model. This approach significantly reduced computation time while maintaining continuity and structural accuracy across block boundaries.

The reconstruction pipeline ultimately produced a high-precision 3D mesh, an aligned colored point cloud, and associated camera poses. The resulting digital twin accurately reflected the building’s architectural complexity, capturing both macro-scale structural features and fine details such as staircases, corridors, and glass walls. This high-fidelity reconstruction formed the foundation for subsequent wireless propagation simulations.

\subsection{Digital Twin Processing}
\subsubsection{Environment Segmentation}
\begin{figure}[!t]
\centering
\includegraphics[width=3.5in]{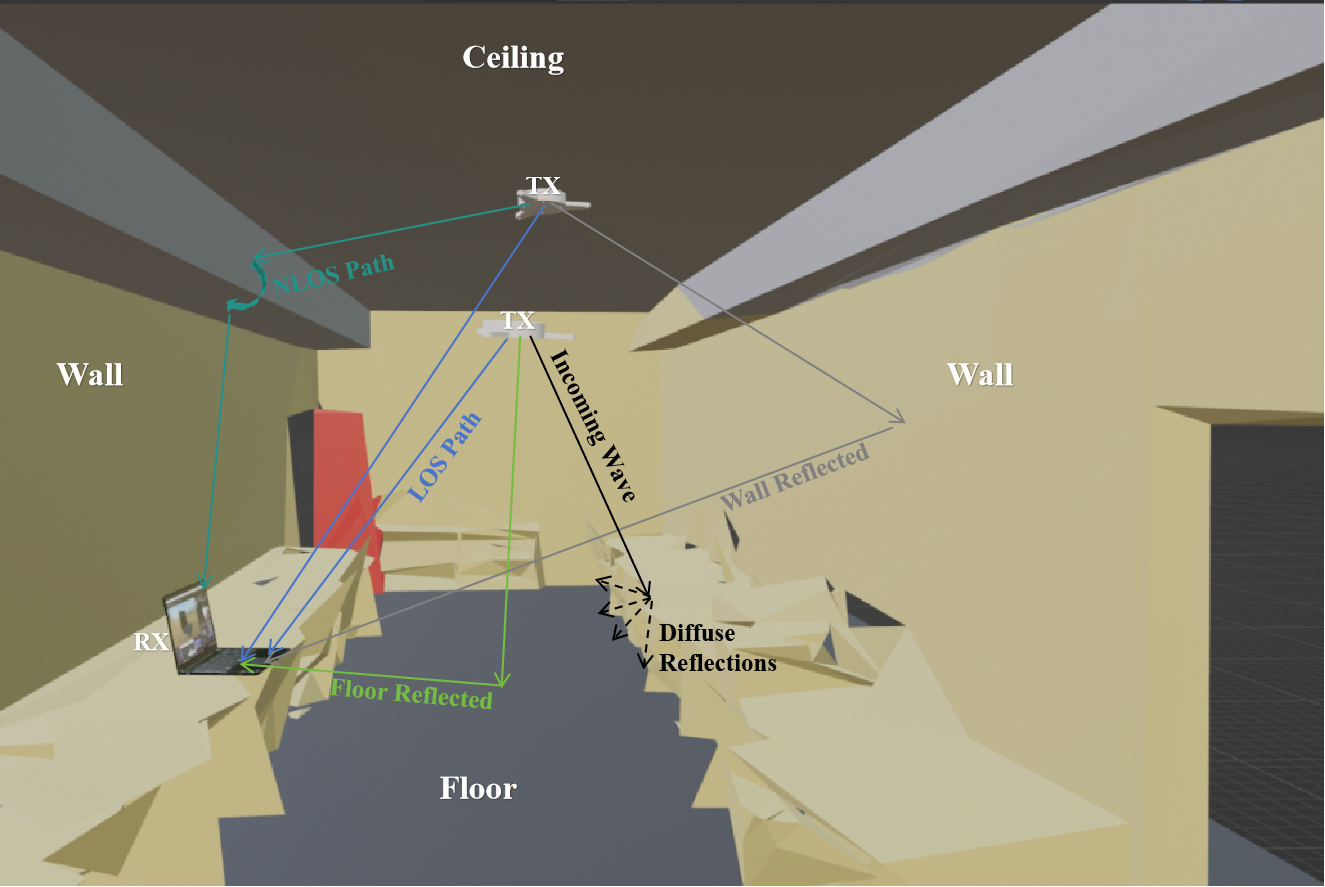}
\caption{Multipath interactions in an indoor wireless scene, including line-of-sight (LOS) transmission, floor and wall reflections, non-line-of-sight (NLOS) diffraction around corners, and diffuse scattering due to surface roughness and irregular objects.}
\label{interaction}
\end{figure}

High-precision 3D modeling provides the geometric foundation for wireless signal propagation simulations. However, accurately modeling signal behavior also requires capturing the electromagnetic characteristics of different materials. In indoor environments, radio waves experience a variety of interactions, including line-of-sight transmission, reflection, penetration, attenuation, and scattering, as illustrated in Fig. \ref{interaction}. First, a direct line-of-sight (LOS) component propagates unobstructed between the transmitter and receiver, often forming the strongest path. When encountering walls, ceilings, or floors, waves may undergo specular reflections, generating additional multipath components with longer delays. At dielectric boundaries such as glass or wooden doors, partial penetration and attenuation occur, altering both the amplitude and phase of the transmitted wave. NLOS links are typically formed by diffraction around corners or sharp edges, which allows energy to bend into shadowed regions. In addition, surface roughness and small irregular objects, such as furniture and electronic devices, introduce diffuse scattering, redistributing energy in multiple directions. 

By segmenting the digital twin into distinct structural elements such as walls, ceilings, floors, doors, and furniture, and assigning them appropriate electromagnetic material properties, the simulation environment explicitly captures these mechanisms to enable accurate modeling of multipath propagation and path diversity.

Conventional 3D scans typically generate continuous surface meshes without distinguishing between materials. For example, a brick wall and an adjacent glass window may be geometrically connected but respond quite differently to electromagnetic waves. Without structural segmentation, it is difficult to reliably identify reflective, penetrable or scatter surfaces, which leads inaccurate of simulation results.

To improve the physical interpretability of the simulation environment, we segment the digital twin model into physically components, including walls, floors, ceilings, doors, windows, tables, electronic devices, and corridors. Each structure is treated as an independent object, allowing for the subsequent assignment of specific electromagnetic material properties. This approach not only enhances the physical realism of the scene but also provides a more accurate foundation for ray-path interaction modeling and propagation parameter calculation.

\subsubsection{Radio Materials Assignment}
After the structural segmentation, electromagnetic material parameters are assigned to each object based on ITU-R P.2040-3 standards. These include the relative permittivity $\epsilon_r$ and electrical conductivity $\sigma$, which define the material’s reflective, penetrative, and scattering behavior during wave propagation. These parameters directly influence path loss and phase response, forming the physical basis for ray tracing and power estimation.

The reflection behavior at dielectric interfaces is modeled using Fresnel reflection coefficients, which take different forms depending on the polarization state. For vertical polarization (i.e., the electric field is perpendicular to the plane of incidence), the reflection coefficient is given by
\begin{equation}
R_\perp = \frac{\eta_2 \cos \theta_i - \eta_1 \cos \theta_t}{\eta_2 \cos \theta_i + \eta_1 \cos \theta_t}
\label{eq:fresnel_perp}
\end{equation}
where $\eta_1$ and $\eta_2$ are the wave impedances of the two media, and $\theta_i$ and $\theta_t$ are the angles of incidence and transmission, respectively. These angles are related by Snell’s law, i.e., $\sin \theta_i / \sin \theta_t = \sqrt{\epsilon_{r2} / \epsilon_{r1}}$.

Different material pairs (e.g., air–glass, air–brick) exhibit significant contrasts in $\epsilon_r$ and $\sigma$, leading to distinct reflection characteristics at their interfaces.

\subsection{Ray-Tracing Based Propagation Modeling}
Once the digital twin is annotated with materials, signal propagation is simulated using deterministic ray tracing, which combines geometrical optics (GO) and the uniform theory of diffraction (UTD). Rays are launched from each transmitter across the 3D angular domain. The total number of rays is determined by the angular sampling density, while the maximum number of interactions per ray is constrained by the allowed propagation depth. These parameters jointly control the trade-off between simulation accuracy and computational efficiency. Rays are terminated early if their power drops below a threshold. Spatial acceleration structures are used to speed up ray–object intersection tests.

At each interaction, the outcome is determined by the material-specific coefficients: Fresnel equations for reflection and transmission, UTD approximations for diffraction, and simplified angular redistribution for scattering. This ensures that the assigned ITU-R parameters directly control how rays evolve in the scene. A valid path is recorded when a ray successfully connects a transmitter and receiver. For each path $P_l$, the delay is
\begin{equation}
\tau_l = \frac{d_l}{c}
\label{eq:delay}
\end{equation}
where $d_l$ is the path length and $c$ is the speed of light. The complex amplitude is the product of interaction coefficients
\begin{equation}
\alpha_l = \prod_{k=1}^{N_l} T_k \, e^{-j\frac{2\pi}{\lambda} d_l}
\end{equation}
where $T_k$ is the coefficient at the $k$-th interaction and $\lambda = c/f$ is the electromagnetic wavelength corresponding to the carrier frequency $f$. The ray tracer outputs a set of multipath components
\begin{equation}
\mathcal{P} = \{(d_l, \tau_l, \alpha_l, \phi_l)\}_{l=1}^L
\end{equation}
where $d_l$, $\tau_l$, $\alpha_l$, and $\phi_l$ represent the path length, delay, amplitude, and phase. All path contributions are aggregated to form the complex baseband CIR at the receiver.
\begin{equation}
h(t) = \sum_{l=1}^{L} \alpha_l e^{j\phi_l} \delta(t - \tau_l)
\label{eq:cir}
\end{equation}
where $L$ is the number of resolved paths, $\alpha_l$ and $\phi_l$ are the amplitude and phase of the $l$-th path, and $\delta(\cdot)$ is the Dirac delta function. 
This CIR is the basic representation form of the channel and also the basis for the calculation of subsequent statistical evaluation indicators.

\section{Experimental Setup}
\subsection{Scenario Description and Parameter Setup}
To evaluate the effectiveness of the proposed ray-tracing-based simulation system, we selected Room 5040 in the SITE building at the University of Ottawa as the evaluation scenario. The office measures approximately 15\,m $\times$ 3.5\,m $\times$ 3.3\,m (L$\times$W$\times$H) and are segmented into six components: walls, door, floor, ceiling, desk, and glass. At a carrier frequency of 2.437 GHz, this study employed two ray-tracing-based wireless simulation platforms: a GPU-accelerated system developed in this work and the commercial software Wireless InSite. Both platforms are based on ray tracing, but they differ in their input formats and internal modeling mechanisms. The proposed system uses XML scene descriptions supported by Mitsuba 3 \cite{Jakob2020DrJit} to define the geometry and material properties of the environment and produces path delays and complex amplitudes that are used to construct the channel impulse response (CIR). On the other hand, Wireless InSite provides a graphical modeling environment and applies a ray-launching and CPU-driven angular grid scanning algorithm to identify multipath components, while it also supports a wider range of standardized propagation models. Because of these differences, it is necessary to adopt a unified configuration strategy to ensure a fair comparison.

\subsection{Simulation Configuration}
Direct parameter alignment between the two platforms is not feasible because of their different ray-tracing mechanisms. To ensure a fair comparison, each simulation was constrained to a fixed runtime of 15 seconds so that both systems execute their ray tracing and channel modeling within the same time budget, rather than reflecting differences in hardware acceleration. All transmitters and receivers were placed in a common Cartesian coordinate system and modeled as isotropic antennas. This setting eliminates variations due to orientation or antenna patterns and ensures that the comparison reflects only propagation effects. A cutoff threshold of –130 dBm was applied, as weaker components lie far below the receiver noise floor and do not contribute meaningfully to channel characterization. The remaining multipath components were ranked by power, and the top 100 are retained for subsequent metric calculation and analysis. This filtering strategy preserves the dominant multipath structure while discarding negligible contributions, thereby balancing computational tractability with modeling fidelity.


\subsection{Evaluation Metrics}

Based on the above configuration, Wi-Fi channel behavior is characterized using a set of performance metrics that capture power, temporal, spatial, and statistical properties of multipath propagation. The path gain of each receiver location (or evaluation cell) is computed as the sum of the powers of all valid multipath components

\begin{equation}
\text{Path Gain} = \sum_{l=1}^{L} |\alpha_l|^2
\label{eq:path_gain}
\end{equation}
Based on the path gain, the received signal strength (RSS) can be calculated by incorporating the transmit power $P_t$ and antenna gains $G_t$ and $G_r$

\begin{equation}
\text{RSS} = P_t \cdot G_t \cdot G_r \cdot \sum_{l=1}^{L} |\alpha_l|^2
\label{eq:rss}
\end{equation}
In multi-transmitter scenarios, the SINR is used to evaluate communication link quality

\begin{equation}
\text{SINR} = \frac{P_r}{\sum_{i \ne t} P_i + N}
\label{eq:sinr}
\end{equation}
where $P_r$ is the received power from the target Tx, $P_i$ denotes the interference powers from other transmitters, and $N = kTB$ is the thermal noise power, with $k$ being Boltzmann's constant, $T$ the system temperature, and $B$ the channel bandwidth.

In addition to power-based metrics, statistical parameters such as the mean delay and root-mean-square (RMS) delay spread are extracted from the CIR to quantify temporal dispersion.
\begin{equation}
\bar{\tau} = \frac{\sum_{l=1}^{L} |\alpha_l|^2 \tau_l}{\sum_{l=1}^{L} |\alpha_l|^2}, \quad
\tau_{\mathrm{RMS}} = \sqrt{\frac{\sum_{l=1}^{L} |\alpha_l|^2 (\tau_l - \bar{\tau})^2}{\sum_{l=1}^{L} |\alpha_l|^2}}
\end{equation}
where $|\alpha_l|^2$ and $\tau_l$ denote the power and delay of the $l$-th multipath component, respectively. The mean delay $\bar{\tau}$ represents the power-weighted average arrival time of multipath signals, whereas the RMS delay spread $\tau_{\mathrm{RMS}}$ measures how dispersed these arrival times are around the mean. A larger $\tau_{\mathrm{RMS}}$ indicates that the multipath components are distributed over a wider time range, which causes the tail of one symbol to extend into the subsequent symbol and results in inter-symbol interference, thereby leading to potential performance degradation.

The spatial dispersion of multipath components is characterized by the Azimuth Spread of Arrival (ASA) and the Zenith Spread of Arrival (ZSA), which are defined as the power-weighted standard deviations of the arrival angles and are widely employed in 3GPP \cite{3GPP} and ITU-R channel models to characterize spatial richness. 

\begin{equation}
\sigma_{\phi} = \sqrt{\frac{\sum_{l=1}^{L} P_l \big( \phi_l - \bar{\phi} \big)^2}{\sum_{l=1}^{L} P_l}}, 
\quad
\bar{\phi} = \frac{\sum_{l=1}^{L} P_l \phi_l}{\sum_{l=1}^{L} P_l}
\label{eq:asa}
\end{equation}

\begin{equation}
\sigma_{\theta} = \sqrt{\frac{\sum_{l=1}^{L} P_l \big( \theta_l - \bar{\theta} \big)^2}{\sum_{l=1}^{L} P_l}}, 
\quad
\bar{\theta} = \frac{\sum_{l=1}^{L} P_l \theta_l}{\sum_{l=1}^{L} P_l}
\label{eq:zsa}
\end{equation}
where $L$ denotes the total number of multipath components, $\phi_l$ and $\theta_l$ are the azimuth and elevation angles of the $L$-th path, $\alpha_l$ is its complex path gain and $P_l = |\alpha_l|^2$ is the corresponding received power. $\phi$ and ${\theta}$ are the power-weighted average azimuth and elevation angles, respectively, while $\sigma_{\phi}$ and $\sigma_{\theta}$ represent the ASA and ZSA, the power-weighted standard deviations of the angles. Large spreads indicate signals arriving from a wide range of directions, which implies richer multipath diversity but also higher interference. Conversely, smaller spreads correspond to more concentrated angular clusters, which are advantageous for beamforming and spatial multiplexing. 

Besides, multipath richness can also quantified through the Ricean K-factor \cite{K-factor}, which measures the relative strength of the dominant line-of-sight (LOS) component against the aggregated non-line-of-sight (NLOS) contributions. Formally, it is defined as

\begin{equation}
K = 10 \log_{10}\left(\frac{P_{\text{LOS}}}{P_{\text{NLOS}}}\right) \ 
\label{eq:kfactor}
\end{equation}
where $P_{\text{LOS}}$ denotes the power of the strongest path and $P_{\text{NLOS}}$ the combined power of all remaining multipath components. A larger $K$ indicates a channel dominated by LOS propagation, while smaller or negative values imply a scattering-rich environment with higher multipath variability.

By jointly considering power, delay, angular-based metrics such as ASA, ZSA and K-factor, the proposed evaluation system achieves a complete characterization of the Wi-Fi channel across temporal, spectral, and spatial domains. In this way, the physically grounded simulation system integrates path-based modeling, metric extraction, and visualization into a unified platform, enabling comprehensive and reproducible analysis of wireless signal behavior in complex indoor environments.

\subsection{Error Metrics and Normalization}
To provide a hardware-agnostic assessment across heterogeneous channel metrics, we utilize two complementary error measures. For linear quantities exhibiting a large dynamic range, we report the relative normalized error with respect to the Wireless InSite reference. This measure allows for consistent comparison regardless of the absolute scale of the metric across different links and scenarios.
\begin{equation}
\varepsilon_{\mathrm{rel}}^{M}(i)
= \frac{\big| M_{\mathrm{GPU}}(i) - M_{\mathrm{WI}}(i) \big|}
       {\big| M_{\mathrm{WI}}(i) \big|}
\end{equation} where $M_{\mathrm{GPU}}(i)$ and $M_{\mathrm{WI}}(i)$ denote the channel metric estimates from our proposed simulator and Wireless InSite respectively for pair $i$. We summarize the scenario level accuracy by computing the mean relative error averaged over all analyzed pairs.

For logarithmic metrics, such as SINR in dB, the conventional percentage error is mathematically ill-conditioned near zero dB and fundamentally lacks physical interpretability in the logarithmic domain. For instance, a small absolute difference near 0 dB would yield an inflated relative error, even if the physical impact is negligible. Therefore, we adopt the absolute deviation as the standardized error metric.
\begin{equation}
\varepsilon_{\mathrm{abs}}^{\mathrm{dB}}(i)
= \big| M_{\mathrm{GPU}}^{\mathrm{dB}}(i) - M_{\mathrm{WI}}^{\mathrm{dB}}(i) \big|
\end{equation} The absolute dB difference is numerically stable, avoids division by small values, and directly reflects the physical degradation in signal quality.

This study employs inferential statistical methods to verify the robustness of the performance enhancements and to exclude the influence of random sampling errors. A 95\% confidence interval is defined to establish a range representing the reliability of the average improvement across different environments. For cases with a small sample size, the Student t distribution is selected rather than a standard normal distribution. This distribution manages the statistical uncertainty in small datasets by providing wider tails than a standard normal distribution. Such a characteristic ensures that the estimates remain valid even when the number of samples is limited.

\begin{equation} \mathrm{CI} = \bar{x} \pm t_{\alpha/2, n-1} \left( \frac{s}{\sqrt{n}} \right) \end{equation} where $\bar{x}$ represents the sample mean of the performance difference while $s$ refers to the sample standard deviation. The variable $n$ denotes the total number of evaluated link pairs and $t_{\alpha/2, n-1}$ represents the critical value for the t distribution at a significance level of 0.05. The lower bound of this interval provides a conservative estimate of the expected gain. This approach prevents an overstatement of the system's advantages. The study also conducts a two-sided paired $t$ test against the null hypothesis of zero improvement. The paired methodology compares the proposed system and the baseline under identical link conditions. This method isolates algorithmic gains from environmental factors such as the path loss variations present in different scenarios. The calculated p-value serves as an objective indicator. It determines whether the improvement in the experiments is a direct result of the algorithm or a lucky coincidence caused by random factors.

\section{Results and Analysis}
This section presents the evaluation results of the proposed measurement system. We first demonstrate the accuracy and fidelity of the reconstructed 3D environment, followed by an analysis of temporal and angular channel characteristics extracted from the CIR. Finally, the multipath richness of the indoor channel is examined using the Ricean K-factor. Together, these results provide a comprehensive assessment of the system’s capability to capture the essential temporal, spatial, and statistical properties of wireless propagation in complex indoor environments.
\subsection{3D Environment Reconstruction}
\begin{figure}[!t]
\centering
\includegraphics[width=3.5in]{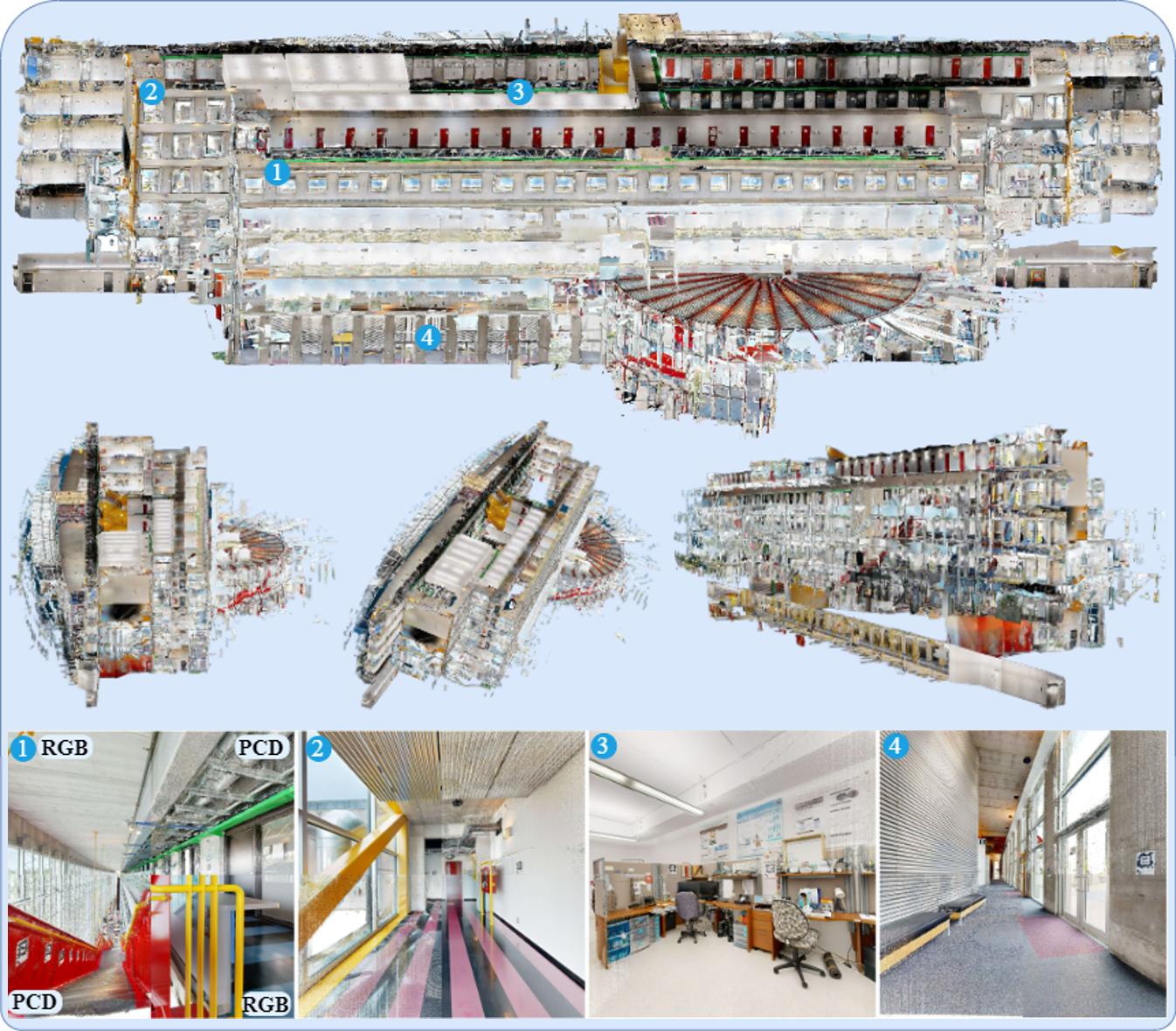}
\caption{Reconstructed 3D digital twin environment: it shows the front view, top view, side view and back view.}
\label{differentview}
\end{figure}

The reconstructed digital twin achieves high geometric accuracy and visual fidelity, as illustrated in Fig.~\ref{differentview}. Multiple views of the 3D model (front, top, side, and back) confirm that the environment preserves structural details such as walls, doors, and ceilings. The alignment process successfully mitigated common issues such as glass reflections and repetitive patterns, ensuring global consistency without notable translation or rotation drift. For instance, the top view of the fifth floor closely matches the official floor plan. The reconstructed mesh further captures challenging elements such as a six-story glass wall, long staircases, and narrow corridors, which demonstrates robustness across complex architectural features. To quantify the geometric uncertainty of the digital twin, we performed a cross-validation analysis by aligning the Matterport Pro3 model against an independent scan acquired via an iPhone LiDAR sensor. The iPhone model served as the reference frame for rigid registration within CloudCompare \cite{cloudcompare} using the Iterative Closest Point algorithm. This process yielded a root mean square alignment error of 0.0118 meters which serves as a quantitative indicator of geometric consistency between the two scans. Although this value appears small, characterizing such alignment uncertainty is critical because geometric deviations directly propagate into the subsequent ray tracing predictions. An RMS position error of approximately 1 cm implies that the reconstructed coordinates of walls and obstacles may be slightly shifted relative to the reference frame. When the ray tracing engine processes this geometry, these spatial shifts translate into corresponding changes in the locations of reflection and diffraction points which ultimately perturb the predicted path lengths and arrival angles of multipath components. In our specific scenario, this centimeter level uncertainty remains significantly smaller than both the typical transmitter-receiver distances and the delay resolution of the system which ensures that its impact on the global channel statistics is limited. We further assessed the cloud to cloud absolute distances after registration and found a mean distance of 0.0408 meters with a standard deviation of 0.0715 meters. These quantitative results demonstrate that the Matterport based digital twin maintains high precision and structural fidelity suitable for large scale indoor propagation modeling. Further details of the comparison setup and processing pipeline can be found in \cite{lidarscanning}.

\subsection{Comparative Evaluation with Wireless InSite}
\subsubsection{Office Scenario}
\begin{figure*}[htbp]
    \centering
    \begin{minipage}[c]{0.8\textwidth} 
        \centering
        \begin{minipage}[b]{0.38\textwidth}
            \centering
            \raisebox{0.4\height}{\includegraphics[width=0.7\linewidth]{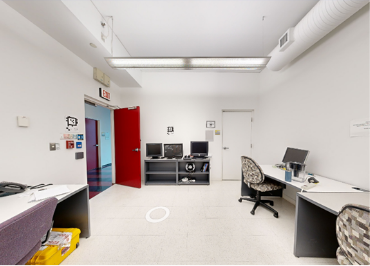}} 
            \subcaption{}\label{fig:a}
        \end{minipage}
        \hfill
        \begin{minipage}[b]{0.58\textwidth}
            \centering
            \includegraphics[width=\linewidth]{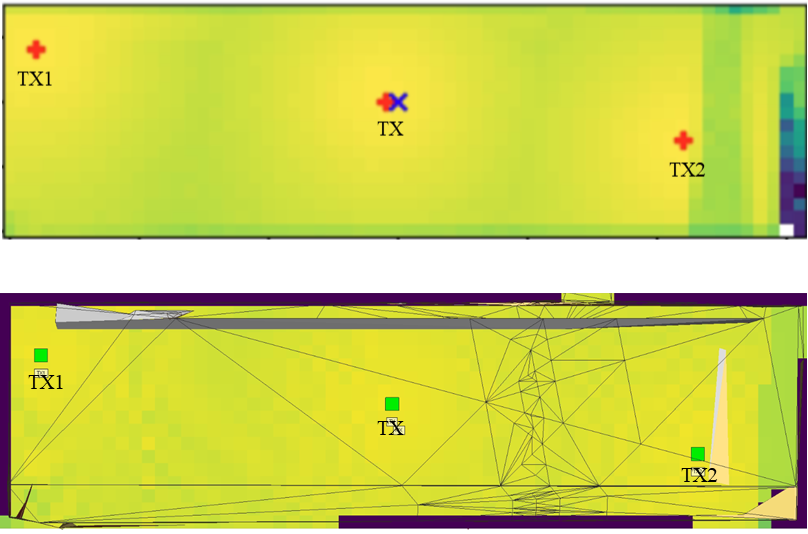}
            \subcaption{}\label{fig:b}
        \end{minipage}
        
        \begin{minipage}[b]{0.38\textwidth}
            \centering
            \vspace{0.2in} 
            \includegraphics[width=0.7\linewidth]{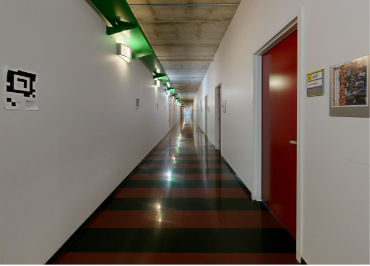}
            \subcaption{}\label{fig:c}
        \end{minipage}
        \hfill
        \begin{minipage}[b]{0.58\textwidth}
            \centering
            \raisebox{0.03\height}
        {\includegraphics[width=\linewidth]{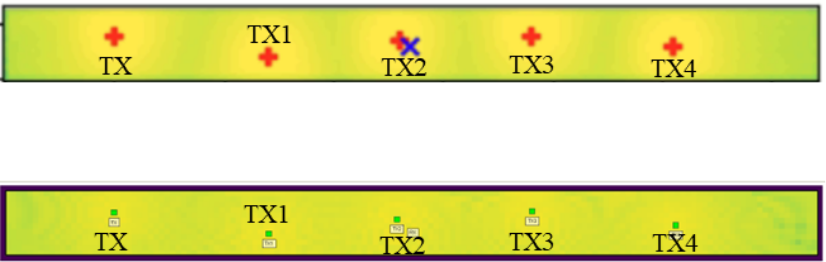}}
            \subcaption{}\label{fig:d}
        \end{minipage}
    \end{minipage}%
    \hfill
    \begin{minipage}[c]{0.18\textwidth}
        \vspace{-0.025\textheight}
        \includegraphics[height=2.89\textwidth]{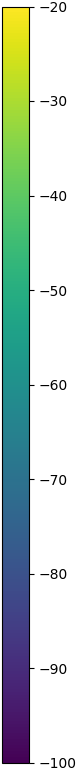}
    \end{minipage}
    
    \caption{Comparison of total received power in the office scenario with 3 transmitters and corridor scenario with 5 transmitters: (a) and (c) are the captured office and corridor environment, (b) and (d) are the corresponding coverage maps generated by the proposed system (top) and Wireless InSite (bottom).}
    \label{room-heatmap-compare}
\end{figure*}

As shown in the coverage maps of Fig. \ref{room-heatmap-compare}(b), the three transmitter positions (Tx, Tx1, and Tx2) are explicitly marked in the layout of the office. Its left subfigure  Fig. \ref{room-heatmap-compare}(b) shows the snapshot of the office environment.

Table~\ref{room-corridor-tableresult}(a) summarizes the results of key metrics. For the Tx-Rx pair, the presented system achieves a mean delay of 10.04~ns and an RMS delay spread of 5.75~ns, compared with 11.97~ns and 9.56~ns from InSite, respectively. This indicates that the developed system captures a more concentrated set of early-arriving paths, preserving more energy in the main component. In terms of path gain, our approach achieves $-38.25$~dB, while InSite gives $-48.81$~dB, resulting in a difference of 10~dB. Although both platforms identify the main path, our method retains more of its energy. SINR values are close in this configuration, as the dominant signal is strong and interference is minimal.

\begin{table*}[t]
\centering
\caption{Comparative evaluation of key channel metrics obtained from the developed ray-tracing system and Remcom Wireless InSite under a fixed simulation time of 15s.}
\label{room-corridor-tableresult}
\renewcommand{\arraystretch}{1.5}

\begin{subtable}{\textwidth}
\centering
\caption{Office scenario.}
\resizebox{\textwidth}{!}{
\begin{tabular}{|c|c|c|c|c|c|c|c|c|}
\hline
\textbf{Platforms} & \textbf{Reflections} & \textbf{Ray samples} & \textbf{Transmitter–Receiver Pair} & \textbf{Mean Delay (ns)} & \textbf{RMS Delay (ns)} & \textbf{Path Gain (dB)} & \textbf{Path Loss (dB)} & \textbf{SINR (dB)} \\
\hline
\multirow{3}{*}{Wireless Insite} & \multirow{3}{*}{6} & \multirow{3}{*}{Ray tracing: \boldmath$2.5^\circ = 10,512$}
& Tx-Rx(central)   & \textbf{11.97} & 9.56 & -48.81 & 48.81 & \textbf{3.80} \\ \cline{4-9}
& & & Tx1-Rx  & 27.30 & 8.17 & -44.77 & 44.77 & -8.40 \\ \cline{4-9}
& & & Tx2-Rx  & 25.95 & 9.52 & -57.57 & 57.57 & -6.95 \\
\hline
\multirow{3}{*}{Proposed System} & \multirow{3}{*}{6} & \multirow{3}{*}{\boldmath$1\times 10^{5} = 100,000$} 
& Tx-Rx(central)   & \textbf{10.04} & 5.75 & -38.25 & 38.25 & \textbf{2.83} \\ \cline{4-9}
& & & Tx1-Rx  & 15.55 & 6.75 & -44.77 & 44.77 & -7.65 \\ \cline{4-9}
& & & Tx2-Rx  & 16.88 & 7.74 & -43.50 & 43.50 & -6.12 \\
\hline
\end{tabular}
}
\end{subtable}

\vspace{1em}

\begin{subtable}{\textwidth}
\centering
\caption{Corridor scenario.}
\resizebox{\textwidth}{!}{
\begin{tabular}{|c|c|c|c|c|c|c|c|c|}
\hline
\textbf{Platforms} & \textbf{Reflections} & \textbf{Ray samples} & \textbf{Transmitter–Receiver Pair} & \textbf{Mean Delay (ns)} & \textbf{RMS Delay (ns)} & \textbf{Path Gain (dB)} & \textbf{Path Loss (dB)} & \textbf{SINR (dB)} \\
\hline
\multirow{5}{*}{Wireless Insite} & \multirow{5}{*}{6} & \multirow{5}{*}{Ray tracing: \boldmath$0.7^\circ = 132,000$} 
& Tx-Rx   & 56.79 & 26.06 & -68.56 & 68.56 & -11.11 \\ \cline{4-9}
& & & Tx1-Rx  & 32.90 & 28.67 & -48.72 & 48.72 & -7.94 \\ \cline{4-9}
& & & Tx2-Rx(central)  & \textbf{12.37} & 18.11 & -47.78 & 47.78 & \textbf{0.39} \\ \cline{4-9}
& & & Tx3-Rx  & 28.36 & 27.18 & -54.50 & 54.50 & -6.66 \\ \cline{4-9}
& & & Tx4-Rx  & 51.15 & 26.74 & -64.11 & 64.11 & -10.60 \\
\hline
\multirow{5}{*}{Proposed System} & \multirow{5}{*}{7} & \multirow{5}{*}{\boldmath$1\times 10^{6} = 1,000,000$} 
& Tx-Rx   & 35.96 & 21.28 & -47.65 & 47.65 & -11.85 \\ \cline{4-9}
& & & Tx1-Rx  & 36.42 & 20.27 & -44.85 & 44.85 & -9.03 \\ \cline{4-9}
& & & Tx2-Rx(central)  & \textbf{16.58} & 10.65 & -37.80 & 37.80 & \textbf{1.62} \\ \cline{4-9}
& & & Tx3-Rx  & 29.97 & 17.67 & -43.42 & 43.42 & -7.12 \\ \cline{4-9}
& & & Tx4-Rx  & 32.43 & 21.72 & -47.23 & 47.23 & -11.41 \\
\hline
\end{tabular}
}
\end{subtable}
\end{table*}

In the Tx1-Rx pair, the transmitter is located in the top-left corner of the office, and the signal must reflect off walls or edges to reach the receiver at the center. Here, both platforms report the same path gain ($-44.77$~dB), indicating that the strongest reflected path was detected in both pairs. However, InSite shows a much higher mean delay (27.30~ns) compared to proposed system (15.55~ns), and a larger RMS delay spread (8.17~ns vs. 6.75~ns), suggesting that InSite's paths are longer and more dispersed. The proposed system extracts shorter and more time-concentrated paths. Regarding SINR, our approach achieves $-7.65$~dB while InSite reports $-8.40$~dB, reflecting an improved ability to distinguish the main signal from interference in a multipath environment.

In the Tx2-Rx pair, the transmitter is located in the bottom-right corner of the office, making this the most challenging setup. There is no direct path to the receiver, and the signal must arrive through multiple reflections.The developed system reports a path gain of 43.50~dB, whereas InSite yields 57.57~dB, corresponding to a difference of more than 14~dB. Our proposed system also shows significantly lower mean and RMS delays (16.88~ns and 7.74~ns vs. 25.95~ns and 9.52~ns), indicating that it successfully extracts shorter and stronger paths even in non-LoS conditions. The SINR difference is smaller, i.e., the proposed system achieves $-6.12$~dB and InSite $-6.95$~dB. Although the gap is less than 1~dB, system proposed  maintains a higher main signal ratio and performs more consistently under interference.

The received power distributions in Fig.~\ref{room-heatmap-compare}(b) visually reinforce these findings. The proposed approach  exhibits concentrated high-power regions from -20 dBm to -40 dBm, especially near the receiver, reflecting its effective capture of main path energy. InSite shows more dispersed low-power areas from -60 dBm to -100 dBm, indicating uneven energy distribution. The data in the Table \ref{room-corridor-tableresult}(a) aligns with the heat map's dispersed trends. Overall, within the 15-second simulation time, this study  outperforms InSite in path density and modeling fidelity.

The angular spread results in Fig.~\ref{fig:aoa-comparison}(a) provide additional insight into these office configurations. For the central transmitter (Tx), both the azimuth spread of arrival (ASA) and zenith spread of arrival (ZSA) are the largest, with ASA around $21^{\circ}$ in proposed system and $30.8^{\circ}$ in InSite, and ZSA is $46.4^{\circ}$ and $52.4^{\circ}$, respectively. These values indicate that multipath energy is dispersed across a wide range of directions, consistent with the dense scattering observed in the coverage map and the relatively lower RMS delay spread. By contrast, Tx1 and Tx2 at the corners yield much smaller spreads (ASA around $26^{\circ}$ and $31^{\circ}$, ZSA below $12^{\circ}$), showing that received energy is concentrated along only a few dominant reflections. These confined angular patterns align with the longer delays reported in Table~\ref{room-corridor-tableresult}(a), reinforcing that transmitter placement in a cluttered indoor environment strongly shapes both temporal and spatial diversity. Between platforms, InSite generally produces wider angular spread, reflecting intention of more delayed and dispersed multipath components. In contrast, our proposed system exhibits tighter angular spread, emphasizing earlier arriving and stronger components. Nevertheless, the overall differences remain modest, and the developed system achieves performance comparable to the commercial simulator.

Table \ref{tab:kfactor} presents the Ricean K-factor results. For the central transmitter position (TX), the proposed system achieves a K-factor of 7.52dB, significantly higher than the 5.27dB reported by InSite. This indicates that the LOS component is stronger and the received energy more concentrated. At the corner transmitters (TX1 and TX2), the K-factor value drop to around 1dB, which reflects the dominance of multipath scattering. 

\begin{table}[htbp]
\centering
\caption{Comparison of Ricean K-factor proposed system and Wireless InSite under different transmitter positions.}
\label{tab:kfactor}
\renewcommand{\arraystretch}{1.1}
\scriptsize
\begin{tabular}{|c|c|c|c|}
\hline
\textbf{Scenario} & \textbf{Transmitter-Receiver Pair} & \textbf{Proposed System (dB)} & \textbf{InSite (dB)} \\
\hline
\multirow{3}{*}{Office} 
 & Tx-Rx (central)   & \textbf{7.52}  & \textbf{5.27}  \\ \cline{2-4}
 & Tx1-Rx            & 1.29  & 1.30  \\ \cline{2-4}
 & Tx2-Rx            & 5.14  & 1.55  \\ \hline
\multirow{5}{*}{Corridor} 
 & Tx-Rx             & -3.78 & -4.59 \\ \cline{2-4}
 & Tx1-Rx            & 1.78  & 1.38  \\ \cline{2-4}
 & Tx2-Rx (central)  & \textbf{8.37}  & \textbf{7.57}  \\ \cline{2-4}
 & Tx3-Rx            & 1.27  & 0.73  \\ \cline{2-4}
 & Tx4-Rx            & -2.94 & -3.76 \\ \hline
\end{tabular}
\end{table}

\subsubsection{Corridor Scenario}
To verify whether the modeling behaviors observed in the office scenario remain consistent across different spatial structures, we extend the comparison to a corridor setting. Compared to the office, the corridor presents longer propagation distances, stronger directional constraints, and more confined reflection geometry. These conditions challenge the ray tracing engines in different ways and are useful for evaluating their path extraction robustness and delay structure resolution. Under the same 15s simulation time , Wireless InSite generates about 132,000 rays with a 0.7° spacing, while the proposed system produces 1,000,000 rays and supports up to 7 reflections. The results are shown in Table \ref{room-corridor-tableresult}(b). 

\begin{figure}[htbp]
    \centering
    \begin{subfigure}[b]{0.48\textwidth}
        \centering
        \includegraphics[width=\linewidth]{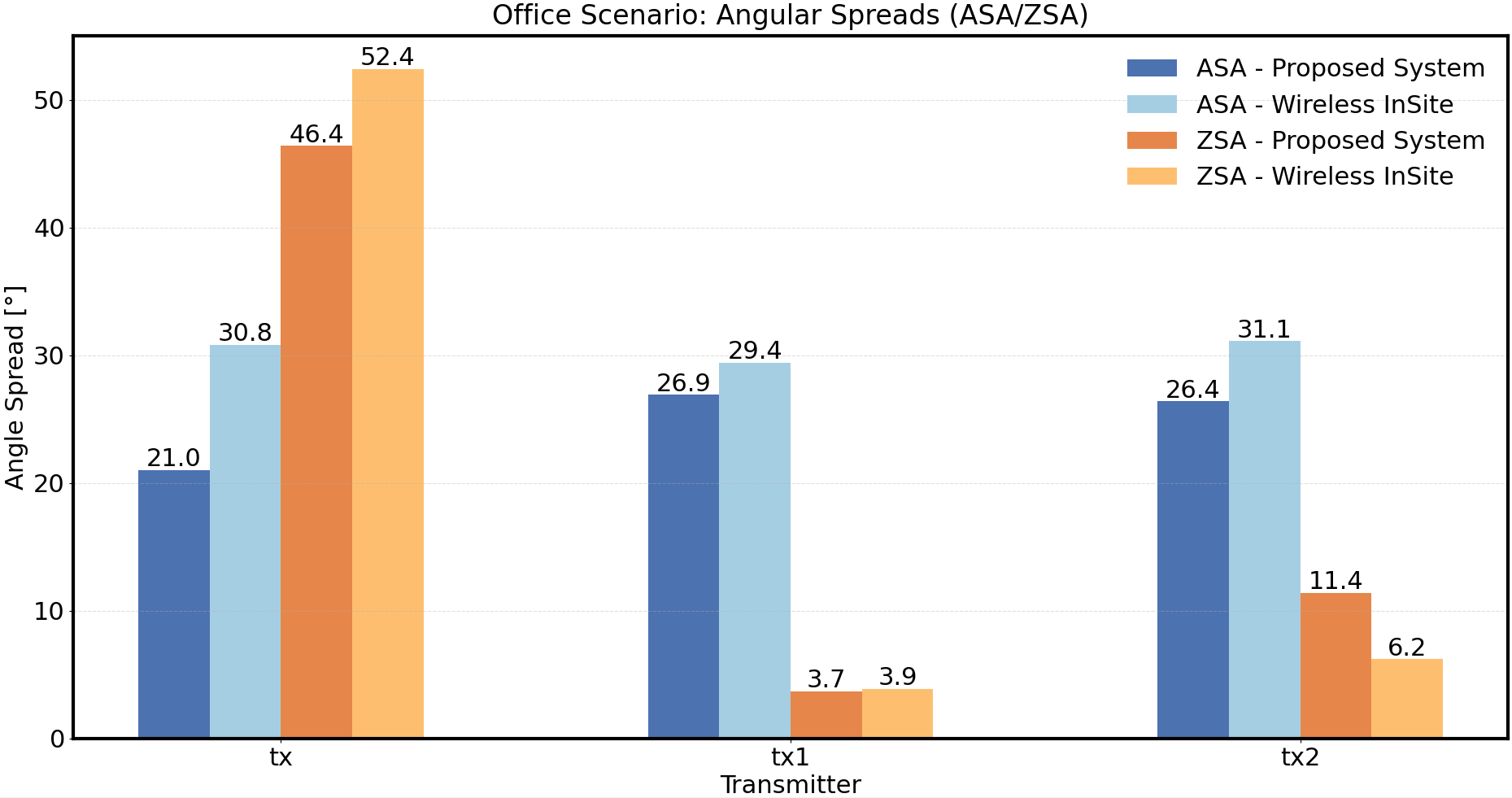}
        \caption{Office scenario.}
        \label{fig:office-aoa}
    \end{subfigure}
    \\
    \vspace{10pt}
    \begin{subfigure}[b]{0.48\textwidth}
        \centering
        \includegraphics[width=\linewidth]{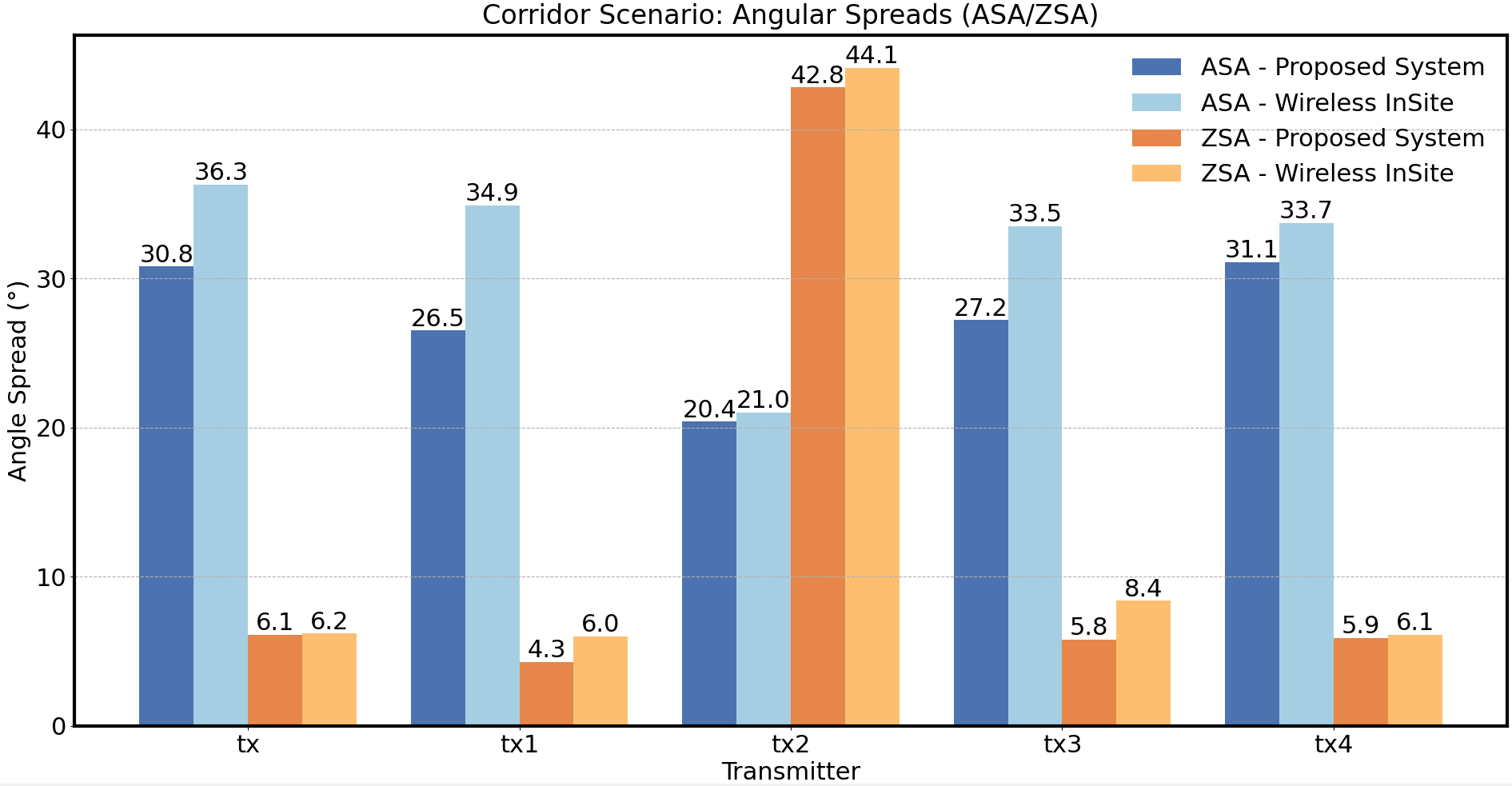}
        \caption{Corridor scenario.}
        \label{fig:corridor-aoa}
    \end{subfigure}
    \caption{Comparison of angular spreads (ASA/ZSA) obtained from proposed system and Wireless InSite in two scenarios: (a) office and (b) corridor.}
    \label{fig:aoa-comparison}
\end{figure}

Fig. \ref{room-heatmap-compare}(d) shows the position of five transmitters and receiver. For the central Tx2-Rx pair, where a direct LOS path exists, the proposed system achieves the strongest performance. The path gain improves to –37.80 dB compared with –47.78 dB in InSite (around 10 dB advantage), with a mean delay of 16.58 ns and RMS spread of 10.65 ns. Although the mean delay is slightly larger than in InSite, the RMS spread is smaller, indicating that multipath energy is more concentrated around the direct path. The corresponding SINR also reaches +1.62 dB, compared with +0.39 dB in InSite, which is consistent with the stronger LOS dominance of the proposed system. These outcomes are consistent with Fig.~\ref{fig:aoa-comparison}(b) where Tx2 exhibits a small ASA ($\approx 20^\circ$) but a very large ZSA ($\approx 43^\circ$), showing that the LOS component dominates in azimuth while strong ceiling/floor reflections broaden the elevation distribution. The K-factor also peaks at 8.37 dB (vs. 7.57 dB in InSite), confirming robust LOS energy preservation.
At the corridor ends (Tx and Tx4), propagation is dominated by long wall-guided reflections. The proposed system still recovers significantly higher path gains (around 15–20 dB advantage) and shorter mean delays, but SINR decreases slightly due to stronger interference. The ASA values here are relatively wide ($\approx 31^\circ$) while ZSA remains narrow ($\approx 6^\circ$), indicating that most energy arrives from lateral reflections constrained to the horizontal plane. This explains both the extended delays and the degraded link quality compared to the LOS center.
For intermediate transmitters (Tx1 and Tx3), the results fall between the two extremes. Path gains improve (by 4–11 dB), RMS delays shrink by 8–10 ns, but SINR again drops modestly (around 1 dB lower than InSite). Angular spreads in Fig.~\ref{fig:aoa-comparison}(b) show moderate ASA ($\approx 26^\circ/27^\circ$) and small ZSA ($< 9^\circ$), reflecting semi-LOS conditions where one or two dominant reflections carry most of the energy, while weaker components overlap and raise the interference floor.

The coverage maps in Fig.~\ref{room-heatmap-compare}(d) illustrate these behaviors visually. The proposed system produces sharper high-power regions (–20 to –40 dBm) around each transmitter, while InSite yields smoother distributions with more stable edge power. This highlights a fundamental trade-off: the proposed engine extracts denser multipath with higher fidelity, which improves path energy and delay accuracy but introduces stronger interference in highly reflective geometries.

Overall, the corridor results confirm that the proposed system consistently enhances path gain by 10–20 dB and reduces delay spread across transmitter positions. While its mean delay is not always smaller, the dispersion of multipath components is generally lower, indicating a more concentrated delay structure. The SINR advantage is most evident under clear LOS (Tx2) but diminishes in strongly reflective NLOS conditions, where richer multipath capture also raises the interference floor. These findings demonstrate that the study provides a more detailed yet interference-aware characterization of indoor channels, which is crucial for reliable digital-twin-based propagation modeling.  

\subsection{Computational Efficiency and Benchmarking}
To ensure hardware-agnostic rigor in the evaluation process, this study introduces a normalized relative error metric to eliminate magnitude differences across various scenarios. The results in Table \ref{acc_summary_compact} demonstrate that the relative error for path loss remains between 15.4\% and 21.2\%, which confirms the stability of the simulator in predicting total signal energy. In contrast, the relative error for time-domain metrics is approximately 30\%. This difference arises from the high sensitivity of delay calculations to late-arriving components in the power delay profile. Specifically, the root mean square delay formula involves squaring the time terms, which causes weak tail signals to amplify calculation biases through this squaring effect. Furthermore, since delay values are measured on a nanosecond scale, even minimal geometric errors in scene modeling can trigger significant percentage fluctuations. Despite these variations in delay metrics, the average deviation of the system-level signal to interference plus noise ratio is restricted to 0.87 dB. This suggests that the 30\% sensitivity error primarily reflects the strict spatial accuracy requirements of time-domain metrics rather than a failure in performance evaluation. Consequently, these findings validate the reliability of the simulator when analyzing complex wireless environments.
\begin{table}[t]
\centering
\caption{Scenario-level prediction accuracy relative to Wireless InSite.}
\label{acc_summary_compact}
\scriptsize
\resizebox{\columnwidth}{!}{%
\begin{tabular}{lccccc c}
\toprule
\multirow{2}{*}{Scenario} &
\multicolumn{3}{c}{Relative error [\%] $\downarrow$} &
\multicolumn{2}{c}{$\Delta$SINR [dB] $\downarrow$} &
\\
\cmidrule(lr){2-4}\cmidrule(lr){5-6}
& Mean delay & RMS delay & Path loss & Mean & Max
\\
\midrule
Office (3 pairs)   & 31.4 & 25.3 & 15.4 & 0.85 & 0.97
\\
Corridor (5 pairs) & 24.7 & 27.6 & 21.2 & 0.87 & 1.23
\\
\bottomrule
\end{tabular}%
}
\end{table}

\begin{table}[t]
\centering
\caption{Accuracy-complexity trade-off of the proposed ray-tracing simulator in the office scenario.}
\label{accuracy_complexity}
\small
\setlength{\tabcolsep}{3pt}
\begin{tabular}{c c c c c c}
\hline
\multirow{2}{*}{Depth} &
\multirow{2}{*}{Rays ($\times 10^{6}$)} &
\multirow{2}{*}{Runtime $t$ (s)} &
\multicolumn{3}{c}{RMS delay} \\
\cline{4-6}
 & & & Mean (ns) & AE (ns) & NE (\%) \\
\hline
 3 & 0.5 &  4.7 & 6.37 & 0.43 & 6.32 \\
 3 & 1.0 &  5.4 & 6.40 & 0.40 & 5.93 \\
 3 & 5.0 &  6.9 & 6.40 & 0.40 & 5.93 \\
 6 & 0.5 & 16.5 & 6.72 & 0.08 & 1.18 \\
 6 & 1.0 & 18.4 & 6.79 & 0.01 & 0.15 \\
 6 & 5.0 & 19.8 & 6.80 & 0.01 & 0.10 \\
 8 & 0.5 & 17.8 & 6.73 & 0.08 & 1.13 \\
 8 & 1.0 & 19.3 & 6.75 & 0.05 & 0.78 \\
 8 & 5.0 & 42.2 & 6.80 & 0.00 & 0.00 \\
\hline
\end{tabular}
\end{table}
To assess computational efficiency independently of specific hardware platforms, we analyze the trade-off between simulation accuracy and complexity. We prioritize the RMS delay spread as the primary metric because it captures the temporal dispersion of energy and remains highly sensitive to multipath components. Table~\ref{accuracy_complexity} presents the runtime and accuracy statistics for the office scenario. To measure algorithmic convergence, we establish the highest complexity configuration with a reflection depth of 8 and 5 million rays as the internal ground truth. The accuracy of other settings is evaluated against this baseline using two metrics. The absolute error measures the deviation in nanoseconds while the normalized error expresses the percentage difference relative to the baseline. The results reveal a distinct efficiency sweet spot where increasing the parameters to 1 million rays and depth 6 causes the normalized error to plummet from 6.32\% to 0.15\%. However, pushing beyond this point to the maximum settings yields diminishing returns where the runtime more than doubles from 18.4 to 42.2 seconds for a negligible accuracy gain of less than 0.15\%. This confirms that the proposed simulator achieves robust accuracy using moderate resources and avoids the need for excessive computational overhead.

\begin{table}[htbp]
\centering
\caption{Quantitative Assessment of Performance Improvement Reliability and Confidence Intervals.}
\label{overall}
\scriptsize 
\setlength{\tabcolsep}{2.5pt}
\renewcommand{\arraystretch}{0.9} 
\begin{tabular}{l ccc cc ccc} 
\toprule
\textbf{Metric} & \multicolumn{3}{c}{\textbf{Rel. Error [\%] $\downarrow$}} & \multicolumn{2}{c}{\textbf{$\Delta$SINR [dB] $\downarrow$}} & \multicolumn{3}{c}{\textbf{Path-gain Improv. [dB] $\uparrow$}} \\
\cmidrule(lr){2-4} \cmidrule(lr){5-6} \cmidrule(lr){7-9} 
& Mean & RMS & PL & Mean & Max & Mean & 95\% CI & p-value \\ 
\midrule
Overall (8 pairs) & 27.2 & 26.7 & 19.0 & 0.86 & 1.23 & 10.9 & [5.3, 16.5] & 0.0025 \\
\bottomrule
\end{tabular}
\end{table}

Table \ref{overall} provides a quantitative assessment of the performance improvement and the statistical reliability across the eight link pairs. The results demonstrate that the developed simulator achieves significant performance gains while maintaining high numerical stability. Specifically, the proposed system exhibits a mean delay relative error of 27.2\%. The specific errors for path loss and root mean square delay are recorded at 19.0\% and 26.7\% respectively. Additionally, the mean absolute deviation for the SINR is merely 0.86 dB. The maximum deviation is strictly limited to 1.23 dB. Such a low deviation level ensures high fidelity in predicting link adaptation performance.
The system achieves a mean path gain improvement of 10.9 dB in terms of performance enhancement. A rigorous statistical analysis was performed to quantify the reliability of this improvement. The two-sided paired t-test yielded a p-value of 0.0025 which is well below the 0.01 significance threshold. This result confirms that the observed performance enhancements are highly statistically significant. Furthermore, the 95\% confidence interval for the path gain improvement ranges from 5.3 dB to 16.5 dB. The gain provided by the system remains substantial even when considering the conservative lower bound estimate of 5.3 dB. This significant gain margin confirms the practical deployment value of the proposed system in heterogeneous channel environments.

\subsection{Validation Against Physical Measurements}

\begin{figure}[!t]
\centering
\includegraphics[width=3.5in]{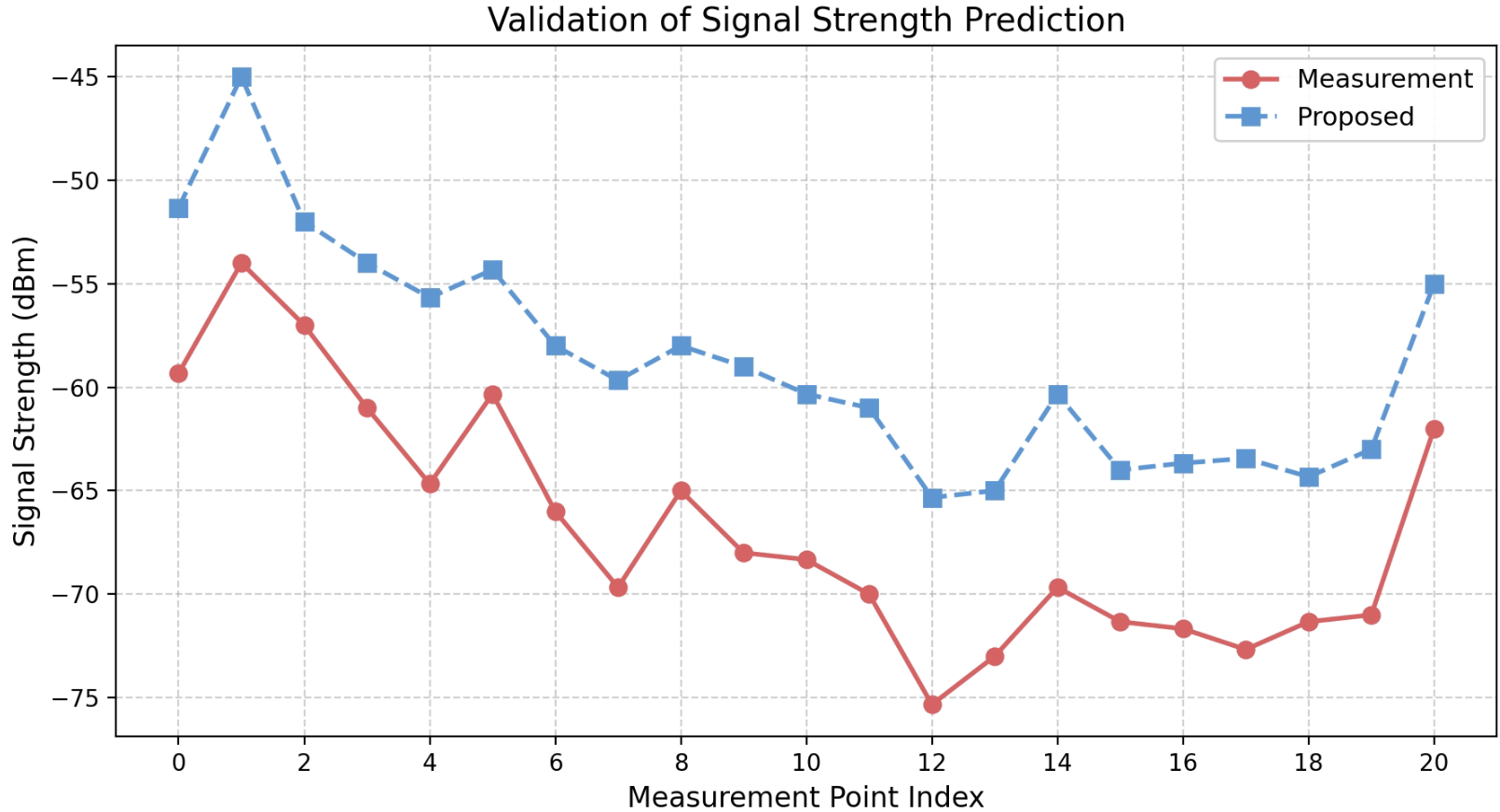}
\caption{Validation of the proposed simulation model against real-world RSSI measurements.}
\label{rssi validation}
\end{figure}

\begin{table}[t]
\centering
\caption{Experimental validation against measured RSSI.}
\label{rssi-validation}
\scriptsize
\resizebox{\columnwidth}{!}{%
\begin{tabular}{lcccc}
\toprule
Method \& Setting          & MAE [dB] & RMSE [dB] & Pearson's $r$ &  offset $b$ [dB] \\
\midrule
Proposed (raw)               & 8.04     & 8.14      & 0.98 & ---  \\
Proposed (Aligned)$^{\mathrm{a}}$   & \textbf{0.97} & \textbf{1.25} & \textbf{0.98} & \textbf{-8.04} \\
\bottomrule
\end{tabular}}
\end{table}

To rigorously validate the fidelity of the proposed simulation model against physical reality, we conducted an onsite measurement campaign. We selected 21 fixed locations uniformly distributed within the target office environment and recorded the empirical Wi-Fi RSSI for the target BSSID at each point, yielding a one-to-one matched set of measured–simulated data pairs. As shown in Fig. \ref{rssi validation}, the proposed simulation model exhibits a fluctuation trend that is highly consistent with real-world measurements, demonstrating a strong spatial correlation. However, a systematic vertical offset is observed. This constant gap is attributed to uncertainties in the actual transmission power and antenna system losses. To isolate the intrinsic modeling accuracy from these constant gain shifts, we applied a level alignment to the simulated RSSI. The quantitative results are summarized in Table \ref{rssi-validation}. Following this calibration, the model achieves high precision with an MAE of $0.97$ dB and an RMSE of $1.25$ dB. The Pearson correlation coefficient $r$ remains at $0.98$, verifying the consistency of spatial variation patterns and confirming that the simulation captures the complex signal propagation characteristics with high fidelity.

\begin{figure}[!t]
\centering
\includegraphics[width=3.5in]{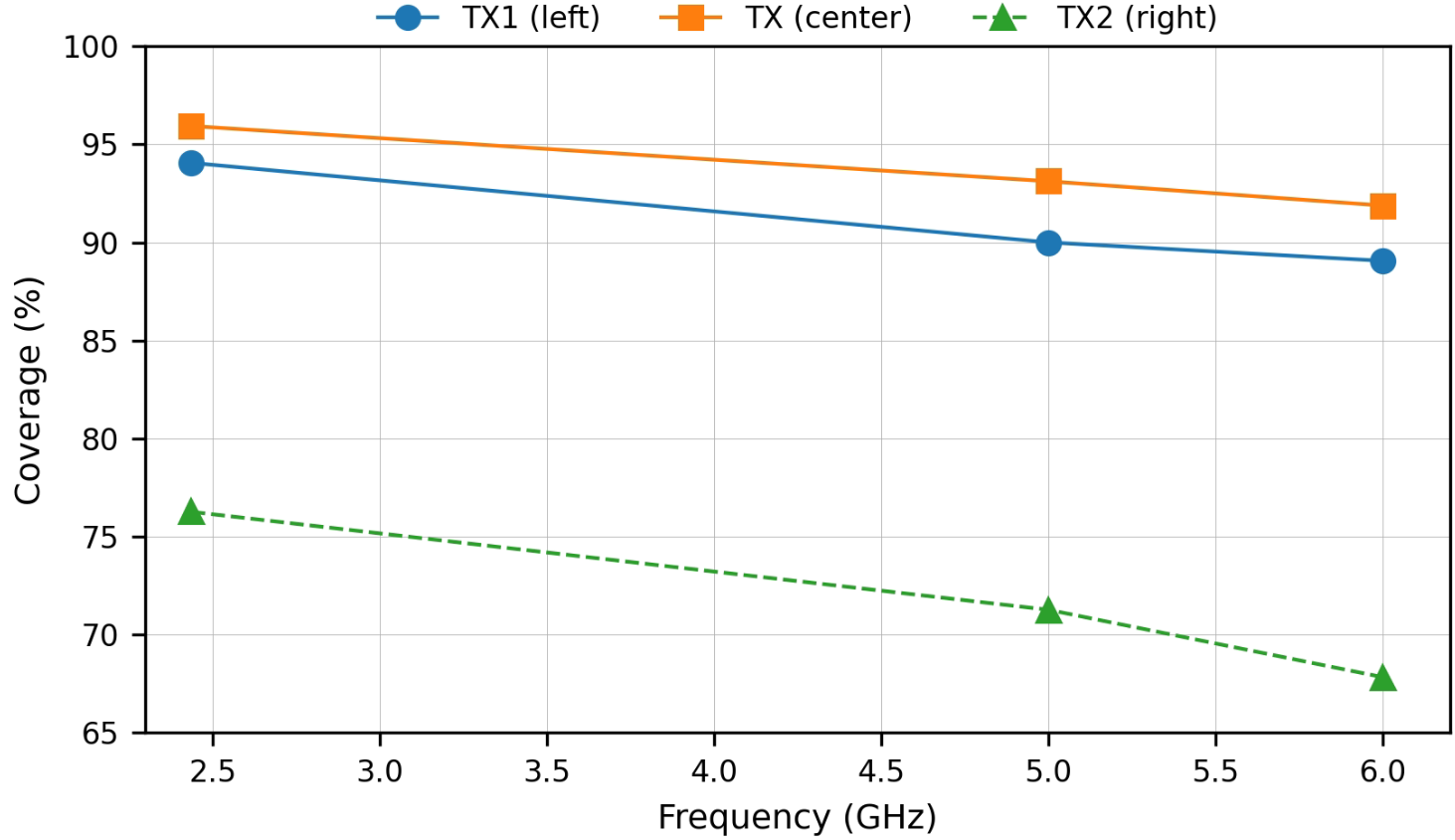}
\caption{Coverage versus carrier frequency for three transmitter placements in the room.}
\label{freq_coverage}
\end{figure}
\subsection{Frequency-Dependent Coverage Analysis}
To investigate the impact of carrier frequency on indoor coverage performance in a controlled setting, we use coverage as the evaluation metric and compute it based on a predefined signal threshold. 
In practical Wi-Fi deployments, a threshold around $-65$~dBm is often used to assess service coverage. However, in our small room scenario, such a threshold would result in nearly full coverage. To emphasize the strong signal region suitable for high data-rate transmission, we therefore adopt a more stringent threshold of $-47$~dBm. Under this fixed configuration, we simulate three carrier frequencies, namely $2.437$~GHz, $5$~GHz, and $6$~GHz, for three different transmitter placements in the room. 
The resulting coverage values at each frequency are reported in Fig.~\ref{freq_coverage}. Crucially, the proposed simulation engine incorporates frequency-dependent material models based on ITU recommendations, where the relative permittivity and conductivity of walls and floors are explicitly modeled as functions of the carrier frequency. When the simulation frequency is changed from $2.437\,\text{GHz}$ to $5\,\text{GHz}$ or $6\,\text{GHz}$, these electromagnetic parameters are automatically updated according to the ITU curves, thereby yielding physically consistent scaling of material attenuation over the $2.4$--$6\,\text{GHz}$ range.

The simulation results show that, for all three transmitter placements, the coverage decreases monotonically as the carrier frequency increases. This trend reflects both the higher free-space path loss at $5\,\text{GHz}$ and $6\,\text{GHz}$ and the stronger material attenuation induced by the frequency-dependent dielectric parameters. The clear frequency-scaling behavior observed here provides indirect validation of the accuracy and effectiveness of our frequency-dependent material modeling, and indicates that the proposed simulation model can reliably characterize indoor propagation in the Wi-Fi~6E/7 bands. These findings further highlight the necessity of explicitly and accurately accounting for carrier-frequency effects when evaluating indoor coverage in digital-twin-based propagation studies, especially for Wi-Fi~6E/7 deployment planning in the $2.4$--$6\,\text{GHz}$ spectrum.

\subsection{Sensitivity and Robustness Analysis}
\subsubsection{Parameters Sensitivity}

\begin{table}[t]
\centering
\caption{Sensitivity of CIR statistics to ray samples and reflection depth.}
\label{sensitivity}
\scriptsize
\begin{tabular}{cccccc}
\toprule
Depth & $N_{\text{rays}}$ 
      & Mean delay [ns] 
      & RMS delay [ns] 
      & Path loss [dB] \\
\midrule
6 & $1\times10^{5}$ & 10.04 & 5.75 & 38.25 \\
6 & $5\times10^{5}$ &  9.79 & 5.73 & 38.53 \\
6 & $1\times10^{6}$ & 10.08 & 5.76 & 38.58 \\
\midrule
3 & $5\times10^{5}$ &  8.50 & 5.72 & 38.94 \\
8 & $5\times10^{5}$ &  9.84 & 5.73 & 38.44 \\
\bottomrule
\end{tabular}
\end{table}
The sensitivity of the CIR statistics regarding the ray budget and the maximum reflection depth in the office scenario is summarized in Table~\ref{sensitivity}. With the reflection depth fixed at six, increasing the number of rays from 100,000 to 1 million causes the mean delay and the RMS delay spread to vary by less than 0.3 ns while the path loss shifts by approximately 0.3 dB. These minor fluctuations indicate that a ray budget of 100,000 already yields stable channel statistics. Conversely when the ray budget is maintained at 500,000 and the maximum reflection depth is extended from 3 to 8, the mean delay increases by approximately 1.3 ns. However the RMS delay spread and the path loss remain virtually constant where the difference between 6 and 8 reflections stays below 0.1 ns for delay and 0.1 dB for path loss. These findings demonstrate that the reported channel metrics exhibit strong robustness with respect to reasonable choices of ray sampling density and reflection depth.

\subsubsection{Materials Sensitivity}
To assess the impact of material property uncertainty arising from diverse indoor materials and layered wall constructions, a sensitivity analysis was performed focusing on the dominant effective wall parameters. Using the ITU-R P.2040-3 standard as a baseline, the relative permittivity and conductivity of the walls were perturbed by $\pm 15\%$ for the representative central transmitter-receiver link under a simulation configuration consistent with Table I. The baseline configuration yielded a path loss of 38.15 dB and a mean delay of 15.27 ns. The simulation results demonstrate a high degree of robustness as increasing the parameters by 15\% resulted in a slight reduction of path loss to 38.09 dB whereas decreasing them led to an increase to 38.22 dB. Consequently, the maximum deviation in path loss was limited to only 0.13 dB and the variation in mean delay remained below 1 ns. These results confirm that the channel impulse response statistics in the considered environment are primarily governed by the macroscopic scene geometry rather than being significantly affected by minor fluctuations in local material properties. This validates that modeling walls with effective homogeneous materials based on ITU standards provides a sufficiently accurate and stable approximation.

\subsection{Discussion}
In addition to these physical-layer results, practical differences between the two platforms are also noteworthy. Wireless InSite automatically exports full simulation directories, which is convenient but can exceed 10 GB per project, whereas the proposed system supports selective data extraction, reducing storage overhead and enabling customized analysis. Moreover, our system provides interactive 3D visualization directly within Jupyter Notebooks, supporting smooth rotation, zooming, and scene exploration, while InSite restricts users to predefined perspectives with limited interaction. These features highlight the lightweight and flexible advantages of the proposed system for both research and practical deployment.

\section{CONCLUSION AND FUTURE WORK} 
This paper introduced a digital twin–based Wi-Fi signal measurement system for indoor wireless channel modeling that integrates LiDAR-based 3D reconstruction, ITU-compliant material assignment, and GPU-accelerated ray tracing. Comparative evaluation with a commercial simulator under equal runtime constraints demonstrates that the proposed system consistently preserves stronger main path energy, produces more concentrated delay profiles, and resolves multipath components with higher fidelity. In line-of-sight scenarios, it provides clearer signal dominance, while in non-line-of-sight cases, it captures richer multipath structures with only minor trade-offs in interference. Crucially, empirical validation against onsite RSSI measurements confirm the system's high fidelity, achieving a spatial correlation of 0.98 and a mean absolute error of 0.97 dB after calibration. Furthermore, the system's capability to model frequency-dependent material attenuation was verified across 2.4 GHz, 5 GHz, and 6 GHz bands, supporting deployment planning for next-generation Wi-Fi 6E/7 networks. These features, combined with the engine's interactivity and scalability, position it as a practical measurement method for digital twin–driven wireless system design and optimization.

While our sensitivity analysis validates the robustness of using effective homogeneous parameters for the current frequency bands, we acknowledge that explicit stratified modeling could offer further fidelity. Future work aims to incorporate multilayer material representations to capture finer propagation details, where surface coatings and internal stratification exert a more significant influence on higher-frequency signal interaction. Additionally, future work will encompass a detailed experimental comparison of our method with other state-of-the-art ray-tracing methodologies,such as those incorporating efficient diffuse scattering models, differentiable ray-tracing techniques, or advanced channel sounding post-processing. This will enable a comprehensive benchmarking of the developed system, further validating its performance and generalizability.

\bibliographystyle{IEEEtran}
\bibliography{ref.bib}

\begin{IEEEbiography}[{\includegraphics[width=1in,height=1.25in,clip,keepaspectratio]{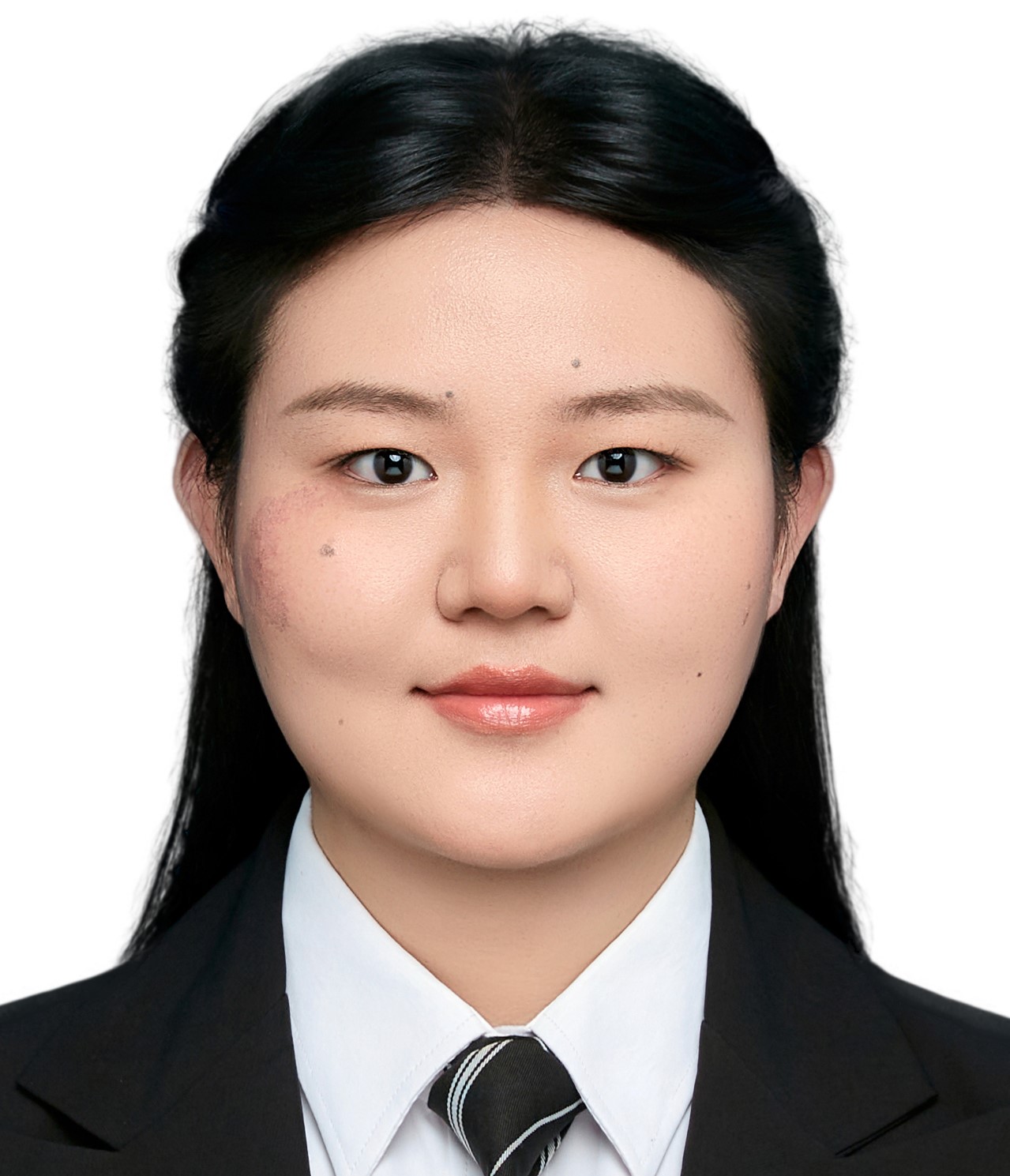}}]{Mengyuan Wang}
(mwang259@uottawa.ca) received the M.Sc. degree in Electrical and Computer Engineering from the University of Ottawa, Canada, in 2024. She is currently pursuing the Ph.D. degree in the School of Electrical Engineering and Computer Science, University of Ottawa. Her research interests include multimedia, metaverse, digital twin, and signal simulation.
\end{IEEEbiography}

\begin{IEEEbiography}[{\includegraphics[width=1in,height=1.25in,clip,keepaspectratio]{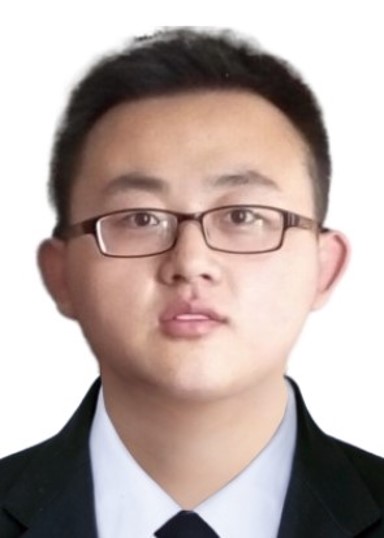}}]{Haopeng Wang}(hwang266@uottawa.ca) received his M.Eng. degree in electronic and communication engineering from Beijing Institute of Technology, Beijing, China, in 2017,  and the Ph.D degree in electrical and computer engineering from the University of Ottawa, in 2024. His research interests are artificial intelligence, extended reality, and multimedia.
\end{IEEEbiography}

\begin{IEEEbiography}[{\includegraphics[width=1in,height=1.25in,clip,keepaspectratio]{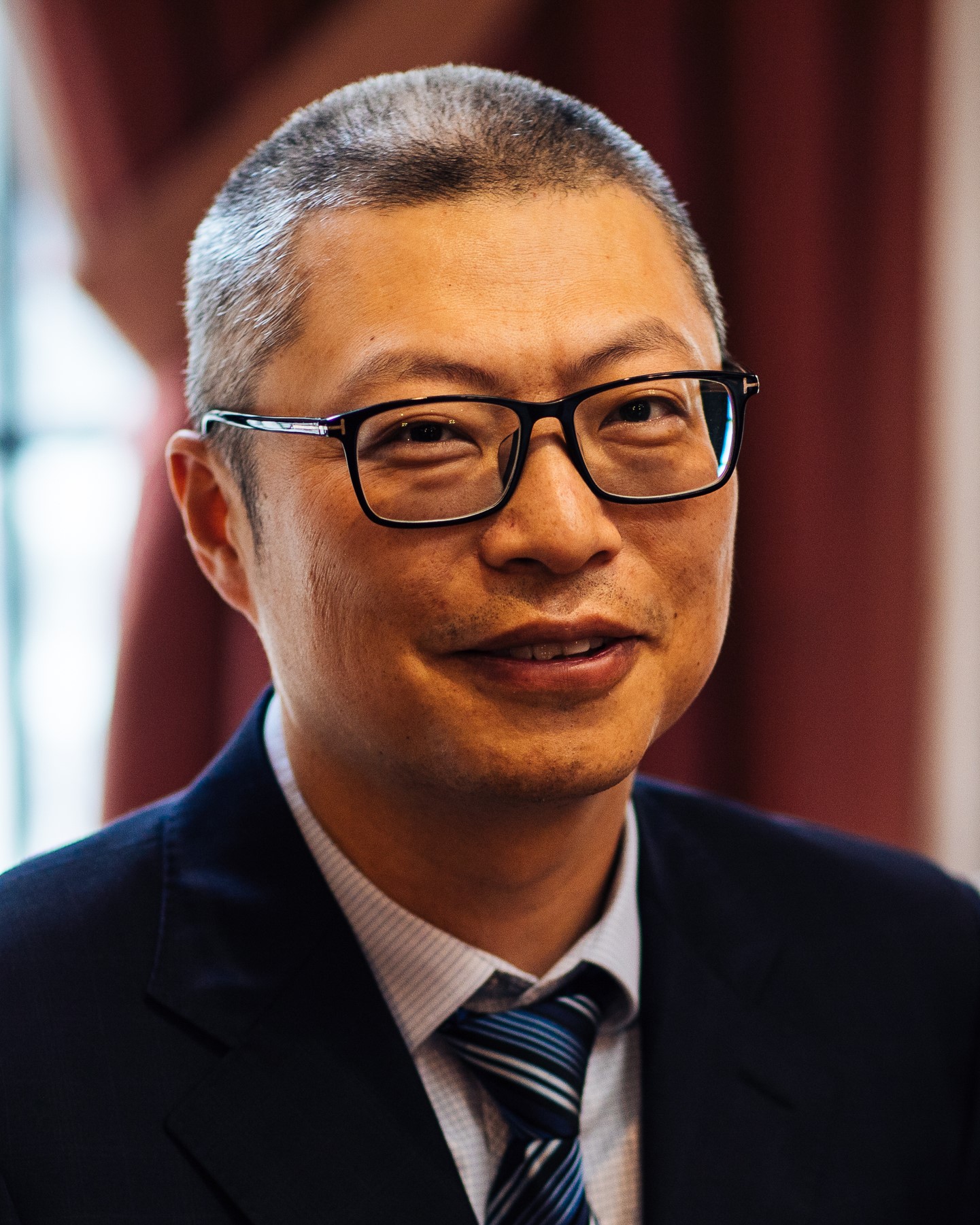}}]{Haiwei Dong} (haiwei.dong@ieee.org) received the M.Eng. degree from Shanghai Jiao Tong University, Shanghai, China, in 2008, and the Ph.D. degree from Kobe University, Kobe, Japan, in 2010. He has held positions as a Principal Engineer at the Artificial Intelligence Competency Center, Huawei Technologies Canada, Toronto, ON, Canada; a Postdoctoral Fellow at New York University, New York City, NY, USA; a Research Associate at the University of Toronto, Toronto, ON, Canada; and a Research Fellow (PD) with the Japan Society for the Promotion of Science, Tokyo, Japan. He is currently an Adjunct Professor at the University of Ottawa, and a Principal Researcher and Director at Huawei Canada, Ottawa, ON, Canada. He is a Senior Member of IEEE and ACM, and a registered Professional Engineer in Ontario. His research interests include artificial intelligence, multimedia, digital twins, metaverse, and robotics. He also serves as a Column Editor of \textsc{IEEE Multimedia Magazine}; an Associate Editor of \textsc{ACM Transactions on Multimedia Computing, Communications, and Applications}; and an Associate Editor of \textsc{IEEE Consumer Electronics Magazine}.
\end{IEEEbiography}

\begin{IEEEbiography}[{\includegraphics[width=1in,height=1.25in,clip,keepaspectratio]{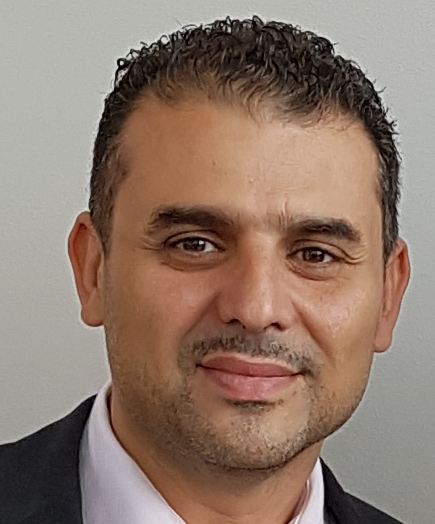}}]{Abdulmotaleb El Saddik}
(elsaddik@uottawa.ca) is currently a Distinguished Professor with the School of Electrical Engineering and Computer Science, University of Ottawa. He has supervised more than 120 researchers. He has coauthored ten books and more than 550 publications and chaired more than 50 conferences and workshops. His research interests include the establishment of digital twins to facilitate the well-being of citizens using AI, the IoT, AR/VR, and 5G to allow people to interact in real time with one another as well as with their smart digital representations. He received research grants and contracts totaling more than \$20 M. He is a Fellow of Royal Society of Canada, a Fellow of IEEE, an ACM Distinguished Scientist and a Fellow of the Engineering Institute of Canada and the Canadian Academy of Engineers. He received several international awards, such as the IEEE I\&M Technical Achievement Award, the IEEE Canada C.C. Gotlieb (Computer) Medal, and the A.G.L. McNaughton Gold Medal for important contributions to the field of computer engineering and science.
\end{IEEEbiography}

\end{document}